\documentclass[sigconf]{acmart}

\usepackage{multirow}

\AtBeginDocument{%
  \providecommand\BibTeX{{%
    \normalfont B\kern-0.5em{\scshape i\kern-0.25em b}\kern-0.8em\TeX}}}

\copyrightyear{2025}
\acmYear{2025}
\acmDOI{10.1145/3765766.3765780}

\setcopyright{rightsretained}

\acmPrice{}

\acmConference[HAI '25] {Proceedings of the 13th International Conference on Human-Agent Interaction}{November 10--13, 2025}{Yokohama, Japan}

\acmBooktitle{Proceedings of the 13th International Conference on Human-Agent Interaction (HAI '25), November 10--13, 2025, Yokohama, Japan}

\acmISBN{979-8-4007-2178-6/25/11}

\usepackage{subcaption}
\usepackage{enumitem}

\usepackage{tikz}
\newcommand\copyrighttext{%
  \footnotesize "© {Owner/Author | ACM} {2025}. This is the author's version of the work. It is posted here for your personal use. Not for redistribution. The definitive Version of Record was published in {Proceedings of the 13th International Conference on Human-Agent Interaction}, http://dx.doi.org/10.1145/3765766.3765780."}
\newcommand\copyrightnotice{%
\begin{tikzpicture}[remember picture,overlay]
\node[anchor=south,yshift=10pt] at (current page.south) {\fbox{\parbox{\dimexpr\textwidth-\fboxsep-\fboxrule\relax}{\copyrighttext}}};
\end{tikzpicture}%
}

\begin{document}

\title{Robots with Attitudes: Influence of LLM-Driven Robot Personalities on Motivation and Performance}


\author{Dennis Becker}
\orcid{0000-0002-1437-6127}
\affiliation{%
  \institution{University of Hamburg}
  \streetaddress{Vogt-Kölln-Straße 30}
  \city{Hamburg}
  \country{Germany}
  \postcode{D-22527}
}
\email{dennis.becker-1@uni-hamburg.de}

\author{Kyra Ahrens}
\orcid{0000-0003-4761-5240}
\affiliation{%
  \institution{University of Hamburg}
  \streetaddress{Vogt-Kölln-Straße 30}
  \city{Hamburg}
  \country{Germany}
  \postcode{D-22527}
}
\email{kyra.ahrens@uni-hamburg.de}

\author{Connor Gäde}
\orcid{0009-0003-2986-6094}
\affiliation{%
  \institution{University of Hamburg}
  \streetaddress{Vogt-Kölln-Straße 30}
  \city{Hamburg}
  \country{Germany}
  \postcode{D-22527}
}
\email{connor.gaede@uni-hamburg.de}

\author{Erik Strahl}
\orcid{0009-0009-7858-8274}
\affiliation{%
  \institution{University of Hamburg}
  \streetaddress{Vogt-Kölln-Straße 30}
  \city{Hamburg}
  \country{Germany}
  \postcode{D-22527}
}
\email{erik.strahl@uni-hamburg.de}

\author{Stefan Wermter}
\orcid{0000-0003-1343-4775}
\affiliation{%
  \institution{University of Hamburg}
  \streetaddress{Vogt-Kölln-Straße 30}
  \city{Hamburg}
  \country{Germany}
  \postcode{D-22527}
}
\email{stefan.wermter@uni-hamburg.de}
\renewcommand{\shortauthors}{Becker et al.}

\begin{abstract}
Large language models enable unscripted conversations while maintaining a consistent personality. One desirable personality trait in cooperative partners, known to improve task performance, is agreeableness. To explore the impact of large language models on personality modeling for robots, as well as the effect of agreeable and non-agreeable personalities in cooperative tasks, we conduct a two-part study. This includes an online pre-study for personality validation and a lab-based main study to evaluate the effects on likability, motivation, and task performance. The results demonstrate that the robot's agreeableness significantly enhances its likability. No significant difference in intrinsic motivation was observed between the two personality types. However, the findings suggest that a robot exhibiting agreeableness and openness to new experiences can enhance task performance. This study highlights the advantages of employing large language models for customized modeling of robot personalities and provides evidence that a carefully chosen agreeable robot personality can positively influence human perceptions and lead to greater success in cooperative scenarios.

\end{abstract}




\maketitle
    \copyrightnotice


\section{Introduction}
\label{sec:introduction}

Human-robot interaction (HRI) is strongly influenced by the perceived robot's personality and has been identified as an important factor for a successful interaction~\cite{Walters2008,Kiesler2002}.
Research has shown that robots are perceived as lifelike and, based on their appearance and behavior, are attributed with a personality~\cite{Meerbeek2008,Kuhne2023}.
Specifically, the perceived personality traits influence the understanding and the anticipation of a robot's behavior~\cite{27ed2bece7d048aa98b1e0b57ed49eed} and attributed personality traits affect the social setting, the perception of the robot, and the perceived quality of the interaction~\cite{Driskell2006}.

\begin{figure}[t]
\centering
    \includegraphics[width=0.5\textwidth]{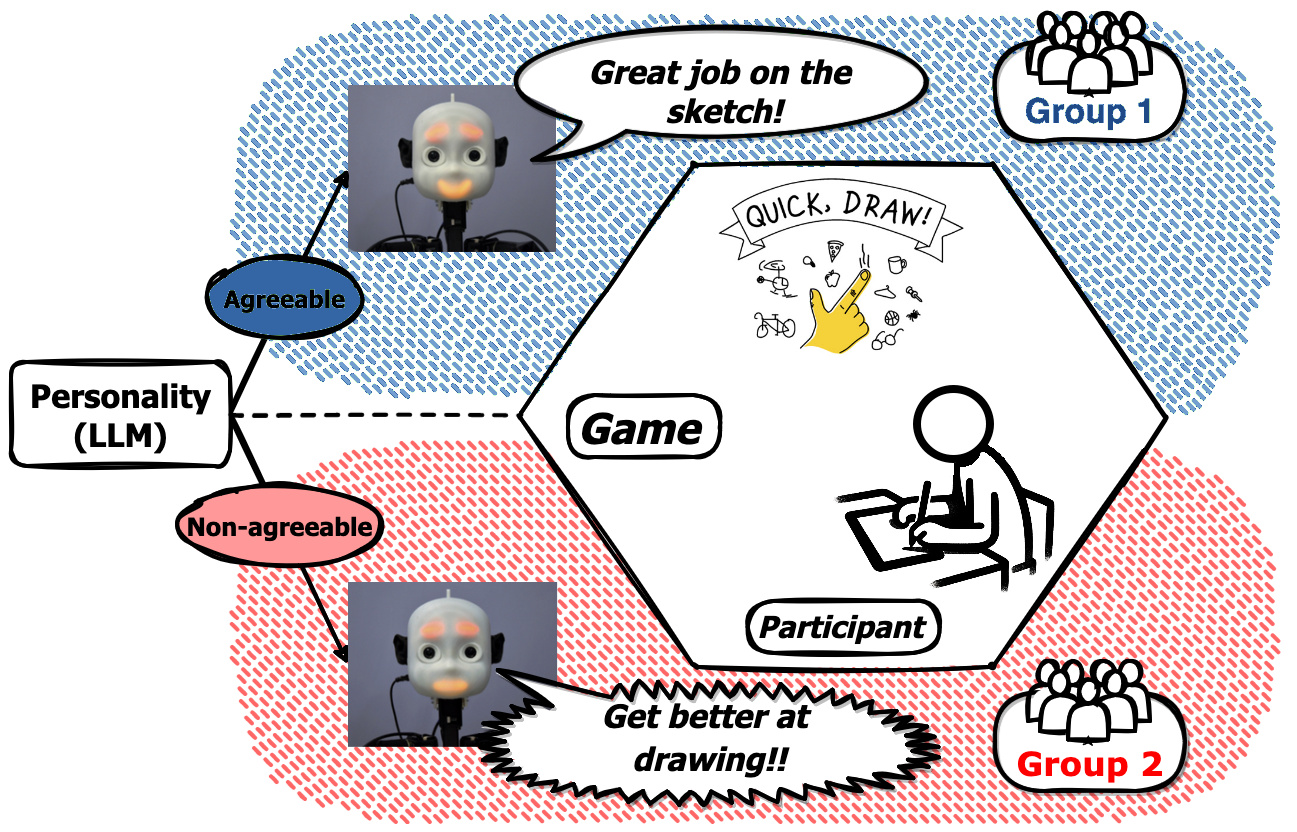}
    \caption{Main study between-subjects design with either an agreeable or non-agreeable robot in the cooperative Quickdraw task. The robot's personality is represented by a large language model that enables unscripted conversation and reactions to the participant's drawing.}
    \label{fig:main_design_img}
\end{figure}

People tend to assign robots personalities and can distinguish between different robots' personalities and distinctive attributes, and recognize their behavior~\cite{Andriella2021}.
A robot's perceived personality can influence various factors, such as likability and acceptance~\cite{Huang2015}.
Moreover, the personality trait of agreeableness is a predictor for team performance and affects cooperation~\cite{Neuman1999,5adce5450c134fa7802d2e4f25bfcb77} and establishment of trust~\cite{Barrick2001}.
A high degree of agreeableness is considered friendly and supportive, whereas low agreeableness is perceived as intolerant and unsympathetic~\cite{Costa2008}.
A cooperation partner with non-agreeable character traits is more likely to make choices and act to prevent the partner from achieving their goals~\cite{Bahamon2017}.

Although the influence of personality on HRI has been investigated, most studies focus on the participant's personality or the robot's personality trait of extraversion and its impact on the interaction while neglecting research on other personality traits~\cite{Esterwood2021b}.
Social robots have the potential to support educational games and develop creativity when paired with humans~\cite{Williams2014}. In a collaborative human-robot scenario, robots can fill the role of providing feedback to promote and encourage creativity~\cite{Lubart2021}, increase self-esteem~\cite{Tang1991}, and motivation~\cite{Mouratidis2008}. For robots to engage in creative and social interactions with humans, they require artificial intelligence, which is facilitated by the present and rapid advancements in the field~\cite{Hoggenmueller2023}.
In earlier work, technical limitations often restricted robot interactions to pre-scripted behaviors specifically designed to exhibit limited personality traits~\cite{RobertJr.2020}. The advent of large language models (LLMs) has opened new possibilities in HRI, especially in areas such as planning~\cite{Singh2023}, social reasoning~\cite{Gandhi2023}, and explainability~\cite{Sobrin-Hidalgo2024}. LLMs enable impromptu, unscripted conversations and the development of more flexible robot personalities~\cite{Wang2024,Zhang2023}.
Additionally, using LLMs in HRI can support a wide range of robot behavior due to an LLM's flexibility through prompt engineering, facilitating a natural interaction.

To overcome previous technical limitations, contribute research on robot personalities beyond extraversion, and advance social robots, this study utilizes the advantages of LLMs to consistently model a robot's personality, with open, unscripted, and individual feedback to investigate the impact of a robot with an agreeable personality on likability, motivation, and task performance. 
Leveraging an LLM with prompt engineering, both agreeable and non-agreeable personalities are modeled and evaluated within the context of a cooperative HRI task. The cooperative task is the game of Quickdraw~\cite{DBLP:journals/corr/HaE17}, a round-based, time-limited drawing and guessing game. During the game rounds, the robot freely expresses its personality, provides participant-specific feedback, and participants can engage in open-ended conversations with the robot between rounds. 
Figure~\ref{fig:main_design_img} provides an overview of the study's between-subjects design.
Our study investigates the effect of a robot's agreeableness in a cooperative task, and addresses three research questions: 
%
\begin{itemize}[leftmargin=0.5in]
    \item[\textbf{RQ1}:] Can an LLM consistently convey a robot's personality?
     \item[\textbf{RQ2}:] Is a robot with an agreeable personality perceived as more likable?
     \item[\textbf{RQ3}:] Does a robot's agreeable personality increase intrinsic motivation and task performance?
\end{itemize}%
%
This paper demonstrates the utility of LLMs for consistent personality modeling and shows that an agreeable robot personality enhances likability and may encourage greater effort, which could result in improved task performance.


\section{Related Work}
\label{sec:related_work}

A robot's personality directly affects the degree of engagement and enjoyableness, due to the social nature of the interactions~\cite{Tay2014}.
For a successful and pleasant interaction, these robots require state-of-the-art technology and social attributes~\cite{Heerink2010}.
Thus, various social characteristics have to be included in the design and implementation of human-robot interaction, such as non-verbal communication and personalities~\cite{Hwang2013,Looije2010}.
Personality can be defined as a set of traits that are expressed as consistent behavioral patterns~\cite[Chapter~9]{Ellis2009} and enables behavioral predictions~\cite{Cattell1950}.
Behavioral patterns are predictive of interaction quality and task performance~\cite{Breazeal2012}, enable understanding of robot actions, and provide a degree of transparency~\cite{Meerbeek2009}.

A frequently used framework for measuring and describing personality traits in human-robot interaction is the Big Five~\cite{Goldberg1990} personality traits. The five personality traits consist of extraversion, conscientiousness, emotional stability, openness to experience, and agreeableness. Extraversion describes the degree of being outgoing, talkative, and sociable, whereas the opposite is described as introversion~\cite{Rhee2013}. Conscientiousness is the extent to which individuals are aware of their actions in terms of mindfulness and organization~\cite{McCrae1992}. Emotional stability is the degree to which an individual remains calm or how easily the person can be upset~\cite{Neuman1999}.
Openness to experience describes creativity, and a person high in that trait can be considered curious and adventurous~\cite{Driskell2006}.
Finally, agreeableness describes a compassionate personality that is characterized by compliance, cooperativeness, and friendliness~\cite{DeYoung2007}.

Research suggests that agreeableness is related to the approaching distance to a robot~\cite{Takayama2009}, a positive perception of a robot
~\cite{Bernotat2017}, and may affect the perceived trust~\cite{Paetzel-Prusmann2021}.
Interacting with robots that exhibit polite and empathetic behavior is generally preferred~\cite{Cramer2010,Nomura2009}, and their positive attitude can increase their perceived anthropomorphism~\cite{Salem2014}, provide benefits in task understanding, and can improve performance~\cite{Leite2008}.
Furthermore, depending on the context, politeness can avoid conflicts~\cite{Salem2013} and positively affect the outcome in a teaching context~\cite{Wang2008}.
On the other hand, rude robots discourage interaction and are perceived as less likable~\cite{Zotowski2015}.
However, research suggests that this general tendency is context- and task-dependent, where for certain tasks impolite strategies can also increase engagement~\cite{Castro-Gonzalez2016,Short2010} and motivation~\cite{Rea2021}.
An additional aspect of the interaction is the robot's performance feedback.
Positive performance feedback and praise increase self-esteem \cite{Ryan2000} and perceived task enjoyment~\cite{Fasola2012}.
In addition, positive feedback can reduce participants' stress and arousal during the interaction~\cite{Marx2024} but may not directly affect intrinsic motivation~\cite{Donnermann2022}.

In a cooperative human-robot task, a robot and a person are in a social collaboration to complete an objective~\cite{Grosz1996}. 
For a successful cooperation and task completion, the robot requires social and emotional intelligence~\cite{Breazeal2012},
which enables the human collaborator to create a conceptual model of the robot that allows reasoning and anticipation of the robot's actions~\cite{Kwon2020}. 
Verbal communication is natural and essential for human-robot interaction, however, it can be challenging due to limited dialogue options and the need for linguistic understanding~\cite{Zhang2023b}. LLMs provide a solution to these problems and enable unscripted and engaging conversation~\cite{Wang2024b} while maintaining a consistent personality~\cite{Galatolo2023,Hilliard2024,Sorokovikova2024}.

Motivation is required to engage and remain focused on collaborative tasks and is fundamental for coordinated and joint actions~\cite{Kathleen2022}.
Various factors can negatively impact human-robot interaction, and humans might be adverse to interacting with robots~\cite{Wang2015}.
Specifically, negative feelings, discomfort, and an expectation gap with the robots' capabilities can negatively affect motivation and engagement~\cite{Kwon2016}.
Furthermore, a lack of trust can lead to resistance to interacting or abandoning the collaborative task~\cite{Lewis2018}.
Emotional and social expressions have been shown to improve communication, and socially engaging robots might further alleviate interaction concerns~\cite{Dautenhahn2007}.


\section{Methodology}
\label{sec:methodology}

\subsection{Research Design} 

To research the validity of adopting an LLM for robot personality modeling, the differences between an agreeable and non-agreeable robot personality, and its effects on performance and intrinsic motivation, an online pre-study for evaluation of the LLM and a lab experiment were designed. An agreeable and a non-agreeable robot personality are implemented using an LLM, and the pre-study estimates whether the difference in agreeableness is successfully conveyed. 
Although LLMs can express personality, research conducted on evaluating personality traits rarely relies on human evaluation~\cite{Yang2024} and suggests that LLM personality traits can be perceived with varying degrees and that smaller models may be inconsistent~\cite{jiang-etal-2024-personallm}. Thus, a separate evaluation of the designed personalities is required. The evaluated personalities are then used for the robot in the main study. Both studies are utilized to estimate if the agreeable personality is correctly perceived in both studies and the effect of a robot's agreeable personality on its likability. Finally, the main study enables an estimation of the effect of the robot's personality on the participants'
motivation and task performance.

For modeling two distinct personalities and enabling an open conversation with the robot, the open-source large language
model Vicuna~\cite{vicuna2023} is used, which enables complete control over the model. Vicuna is based on the Meta LLaMA~\cite{touvron2023llama} model, fine-tuned with user-shared dialogues, and can engage in prolonged multi-turn conversations. The Vicuna model is prompted to act as an art teacher with the addition of the personality. Specifically, the generated responses reflect either an agreeable or non-agreeable personality.

\subsection{Pre-study}

\begin{figure}
\centering
\begin{subfigure}{.5\linewidth}
  \centering
  \includegraphics[width=.995\linewidth]{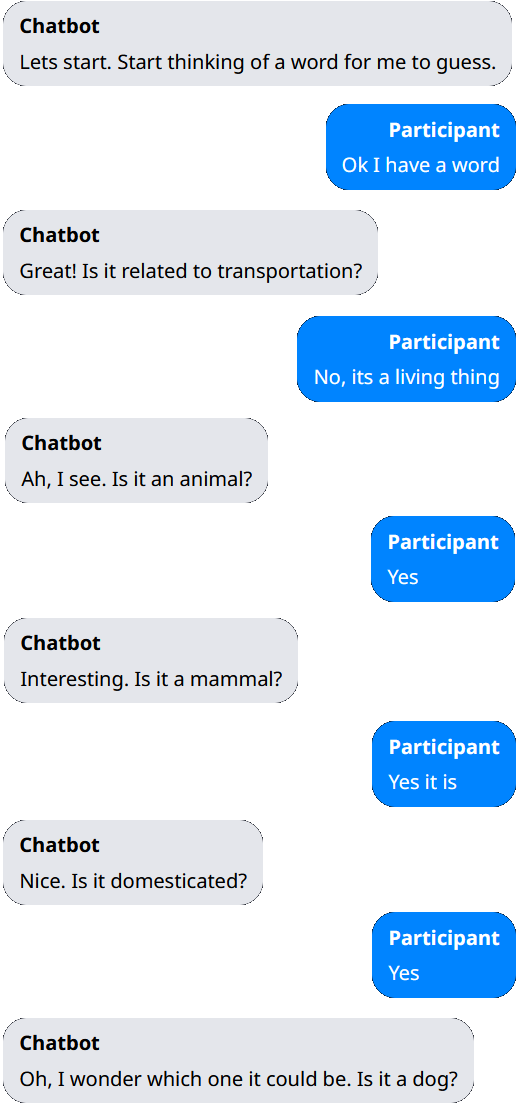}
  \caption{Agreeable chatbot}
  \label{fig:sub1}
\end{subfigure}%
\begin{subfigure}{.5\linewidth}
  \centering
  \includegraphics[width=.995\linewidth]{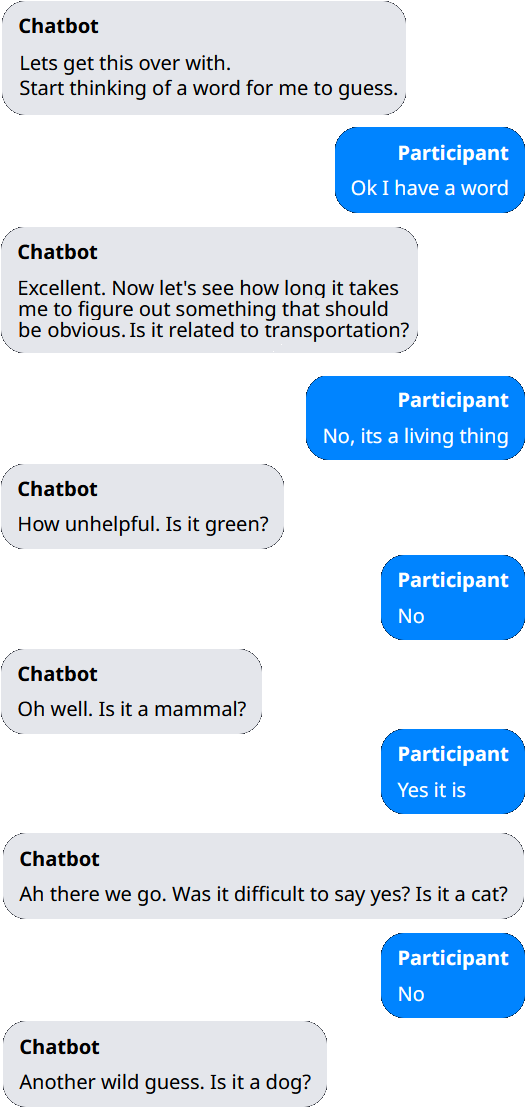}
  \caption{Non-agreeable chatbot}
  \label{fig:sub2}
\end{subfigure}
\caption{Part of the dialogue between a human and a chatbot used in the pre-study. The non-agreeable robot appears less interested in the task and is uncooperative.}
\label{fig:prestudy_inter}
\end{figure}

To evaluate the implemented personalities of the LLM, an online study was conducted.
Two dialogues are created for the pre-study by interacting with the implemented personalities. 
In the online survey, participants are presented with one of the created chat dialogues between a human and the chatbot.
In the interaction, the chatbot has to guess a word that the person in the interaction has in mind. Thus, the chatbot asks questions to narrow down the possibilities, express its personality, and finally guesses the correct word. While the agreeable chatbot politely asks goal-oriented questions, the non-agreeable chatbot often makes snarky comments and expresses little interest in the interaction. This behavior can be noted in the initial dialogue, provided for both groups in Figure~\ref{fig:prestudy_inter}.

The participants are randomly assigned to one of the conditions, agreeable or non-agreeable. After reading the chat interaction between the human and the chatbot, their perception of the chatbot and its perceived personality are assessed.

\subsection{Experiment Design}
\label{subsec:experiment_design}

To evaluate the difference between an agreeable and a non-agreeable robot, in the main study, participants play the game of Quickdraw with a robot. In the between-subjects experiment, the participants are randomly assigned to one of the conditions and briefed to be art students partaking in a workshop session with an art teacher robot that will guess their drawings.
Quickdraw has been used for exploring cooperative drawing~\cite{Jansen2021} and creativity tasks for children~\cite{Ali2020}.
The game Quickdraw is a collaborative task where the participant is asked to draw a picture on a touchscreen within 30 seconds.
The robot's role is to guess the object drawn by the participant. 
While the participant is drawing, the robot attempts to guess the object and comments on the drawing, incorporating its personality.
If the robot guesses the object correctly, the round is successful, and the participant receives a point. Thus, both the robot and the participant have to cooperate to maximize the score.

In total, the experiment consists of 12 rounds, and after each round, the robot comments on the participant's drawing, and the participant can freely chat with the robot.
The game Quickdraw provides a dynamic and engaging social interaction between the participant and the robot, which enables the robot to express its personality.

\subsection{Experiment Setup}
\label{subsec:experiment_setup}

\begin{figure}[t!]
\centering
    \includegraphics[width=0.5\textwidth]{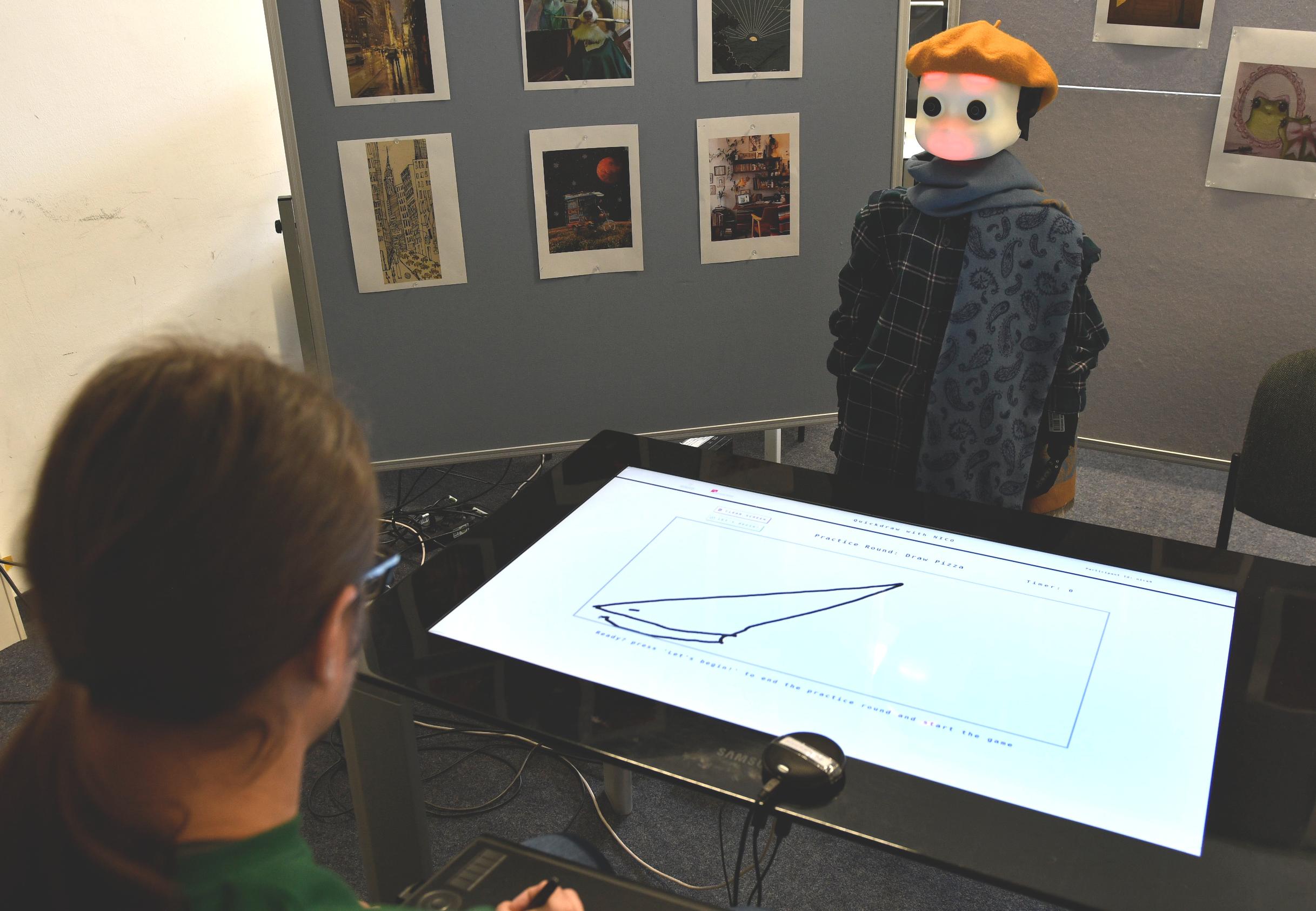}
    \caption{Interaction with an agreeable or non-agreeable robot in a cooperative drawing task. The participants need to sketch an object (in the image, the requested object is a pizza), while the robot comments and attempts to guess the correct object.}
       
    \label{fig:main_study_img}
\end{figure}

For the laboratory experiment, the Neuro-Inspired COmpanion (NICO)~\cite{kerzel2017nico}, a humanoid robot specifically designed for HRI studies, is employed. In the experimental setup, participants are seated in front of a touch table, with the robot positioned on the opposite side. A trackpad is connected to the table, allowing participants to draw using a pen. A custom user interface (UI) is developed specifically for the Quickdraw game to facilitate the interaction. The UI features a drawing canvas, an eraser tool, and a timer indicating the remaining time for each round. Additionally, the UI displays relevant information such as the object to be drawn, the current round number, the participant's score, and instructions such as a prompt to converse with the robot after each round. Prior to the start of the experiment, participants engage in a practice round to become familiar with the trackpad and the UI's functionality. Figure~\ref{fig:main_study_img} illustrates the interaction between a participant and the robot during the experiment.

For the robot's guesses regarding the participant's drawings, a neural network consisting of three repeated blocks of a convolutional and a max pooling layer followed by two final dense layers was trained on the Quickdraw dataset~\cite{DBLP:journals/corr/HaE17}. The neural network receives the drawing as input and provides the probabilities for each object. The five objects with the highest probability are provided to the LLM, which generates the robot's response and guesses while the participant is drawing. After the drawing time ends, the robot guesses the object with the highest probability provided by the neural network.
For nonverbal communications that complement the robot's personality, varying movements and facial expressions are implemented based on the robot's personality~\cite{Urakami2023}.
If the agreeable robot guesses the drawing correctly, it will nod and raise its arms, give a thumbs-up, and display a happy facial expression. If the guess is incorrect, the robot will move its head down and from side to side while displaying a frown.
Similarly, the non-agreeable robot, when the guess is correct, will nod its head while displaying a neutral emotion.
If the guess for the non-agreeable robot is incorrect, the robot will use its hands to facepalm and shake its head with an angry facial expression.

A pilot study with eight participants
was conducted, and after the experiment and assessing the questionnaires, the participants were interviewed.
Based on the evaluated data and the conducted interviews, minor changes to the experiment setup were applied.
Particularly, the latency of the robot's responses was reduced, and the visibility of information provided on the UI was improved.

\subsection{Questionnaires and Measurements}
\label{subsec:measurements}

To evaluate the effect of the robot's personality, questionnaires are assessed, and as objective measures, the interaction time and the game score are recorded. 

For measuring the perception of the robot, the Godspeed questionnaire~\cite{Bartneck2009} is used, which measures the robot's perceived anthropomorphism, animacy, likeability, intelligence, and safety.
The robot's personality is assessed with the Ten-Item Personality Inventory (TIPI)~\cite{Gosling2003} questionnaire, which measures five key personality traits: openness, conscientiousness, extraversion, agreeableness, and emotional stability.
To measure the participants' motivation during the experiment, the Intrinsic Motivation Inventory (IMI)~\cite{Ryan1983} is utilized, which measures interest/enjoyment, perceived competence, perceived choice, pressure/tension, and relatedness.
Additionally, the score for each participant is recorded. The score expresses how many drawings the robot could guess correctly and measures the task performance. Further, the interaction time with the robot is recorded, which indicates engagement with the robot.

\subsection{Procedure}
\label{subsec:measurements_study_design}

First, the participants' informed consent is obtained, and then the demographics are assessed. Afterward, the game rules are introduced and the participants are escorted to the experiment room, where they are seated in front of the touch table and facing the robot.
The experiment begins with the robot explaining the rules, followed by a practice round to familiarize the participant with the touch table and the trackpad.
After completion of all 12 rounds, the robot instructs the participant to return to the previous room and proceed with the post-experiment questionnaire.

\subsection{Participants}
\label{subsec:participants}

After the evaluation of the design of both studies by the local ethics commission, participants were recruited.
For estimation of the pre-study sample size, a moderate effect size~\cite{cohen1988spa} of Wilcoxon $r = 0.42$ is utilized based on reported results on an agreeable and neutral chatbot~\cite{Volkel2021}, with $\alpha = .05$ and a statistical power of $.95$. Simulation of these assumptions yielded an estimated sample size of 32 participants per group.
The pre-study was completed by 66 participants, with 33 participants in the agreeable and 33 participants in the non-agreeable condition.
Of these participants, 34 are female, and the participants' age ranges from 18 to 66 years ($M = 28.28$, $SD = 10.86$).
Most of the participants are working part-time or full-time (73\%) followed by students (24\%) and those who are retired or are currently unemployed (3\%). In addition, 79\% of the participants have previous experience with LLMs, and 53\% used artificial intelligence in a work-related context.

In contrast to the online pre-study, the effect in the main study, where participants interact with the robot, should be stronger. The reported results of robots' agreeableness on acceptance~\cite{Esterwood2021b} suggest a large effect size. We utilize an effect size of Wilcoxon $r = 0.44$, with $\alpha = .05$ and a statistical power of $.95$, which results in an estimated sample size of 29 participants per group.
The main study had 68 participants, but due to technical issues during the experiment, five participants had to be excluded, resulting in a total of 63 participants. In the agreeable robot condition are 30 participants and 33 participants in the non-agreeable condition.
25 participants are female, and the age ranges from 18 to 64 years ($M = 27.45$, $SD = 8.11$).
The majority of participants are students (70\%) followed by those working part-time or full-time (30\%).


\section{Results}
\label{sec:results}

\subsection{Pre-study Results}

\begin{figure}[h!]
\centering
    \includegraphics[width=0.49\textwidth]{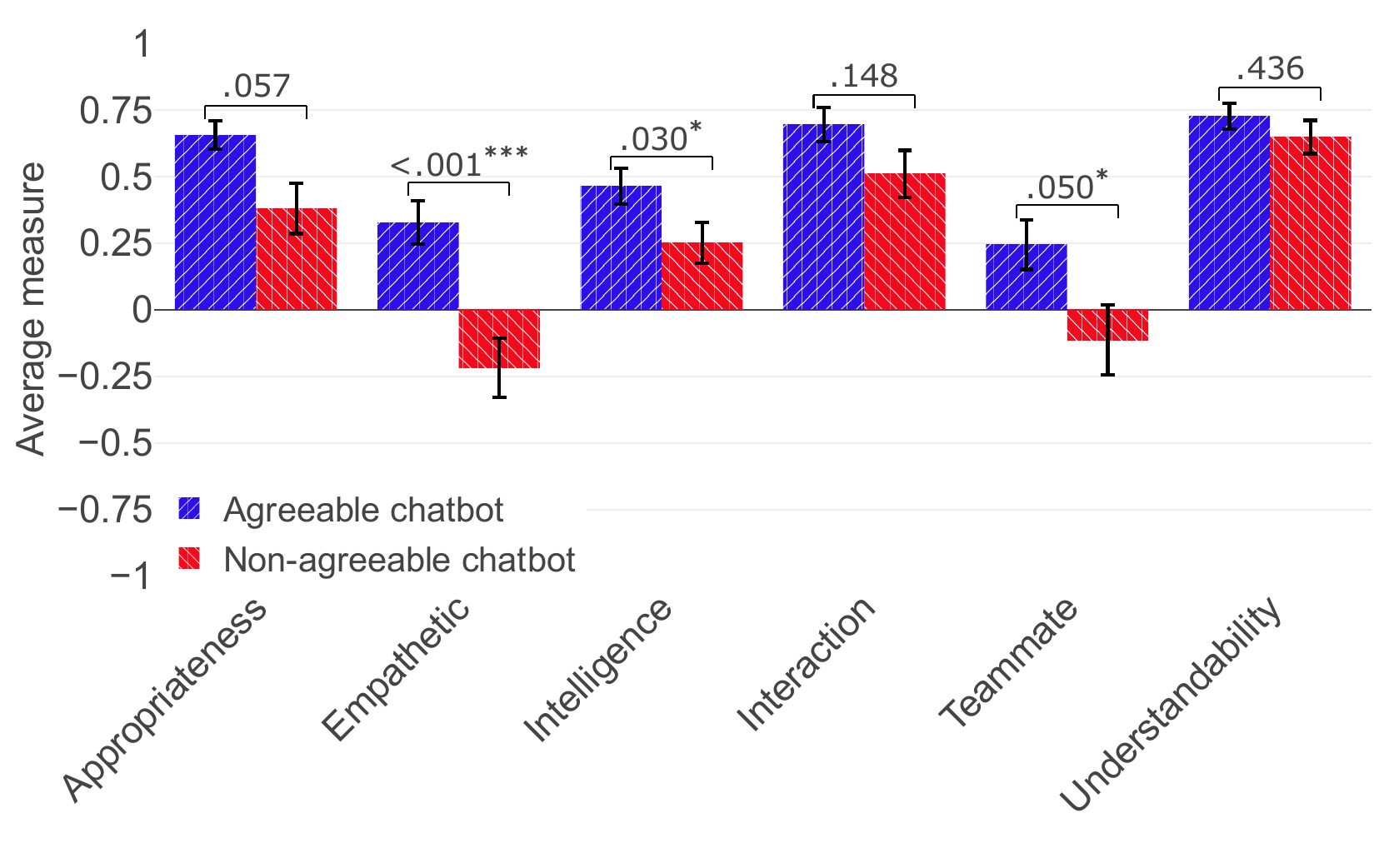}
    \caption{Perception of the chatbots in the pre-study. Values indicate \textit{p} value, \ensuremath{^{*}}\textit{p} $<$ .05, \ensuremath{^{***}}\textit{p} $<$ .001.}
    \label{fig:pre_robot_preceived}
\end{figure}

After the participants read the provided chat interaction between a human and a chatbot, a questionnaire regarding the chatbot and its TIPI was assessed.
While the TIPI is assessed on a Likert scale, as described in Section~\ref{subsec:measurements}, for the perception of the chatbot, the participants rated the chatbot's perceived traits with a slider ranging from strongly disagree to strongly agree. A sentiment of 0 represents a neutral sentiment. The questions are provided in Appendix~\ref{Appendix:chatbotQustionaire}.
An illustration of the perceived difference between both chatbot interactions is shown in Figure~\ref{fig:pre_robot_preceived}.

From the responses, a clear difference between both chatbots can be noticed.
The Mann-Whitney U test suggests a significant difference ($W = 814.5$, $p <.001$) in the chatbots' empathy between the agreeable ($M = 0.33$, $SD = 0.47$) and non-agreeable chatbot ($M = -0.22$, $SD = 0.63$).
Further, a difference between the perceived intelligence of both chatbots is indicated ($W = 713.5$, $p = .030$), where the agreeable chatbot is perceived as more intelligent ($M = 0.47$, $SD = 0.39$) than the non-agreeable chatbot ($M = 0.25$, $SD = 0.44$).
Finally, the participants would significantly ($W = 697.5$, $p = .050$) prefer the agreeable chatbot ($M = 0.25$, $SD = 0.53$) as a teammate over the non-agreeable chatbot ($M = -0.11$, $SD = 0.75$).

\begin{figure}[h!]
\centering
    \includegraphics[width=0.49\textwidth]{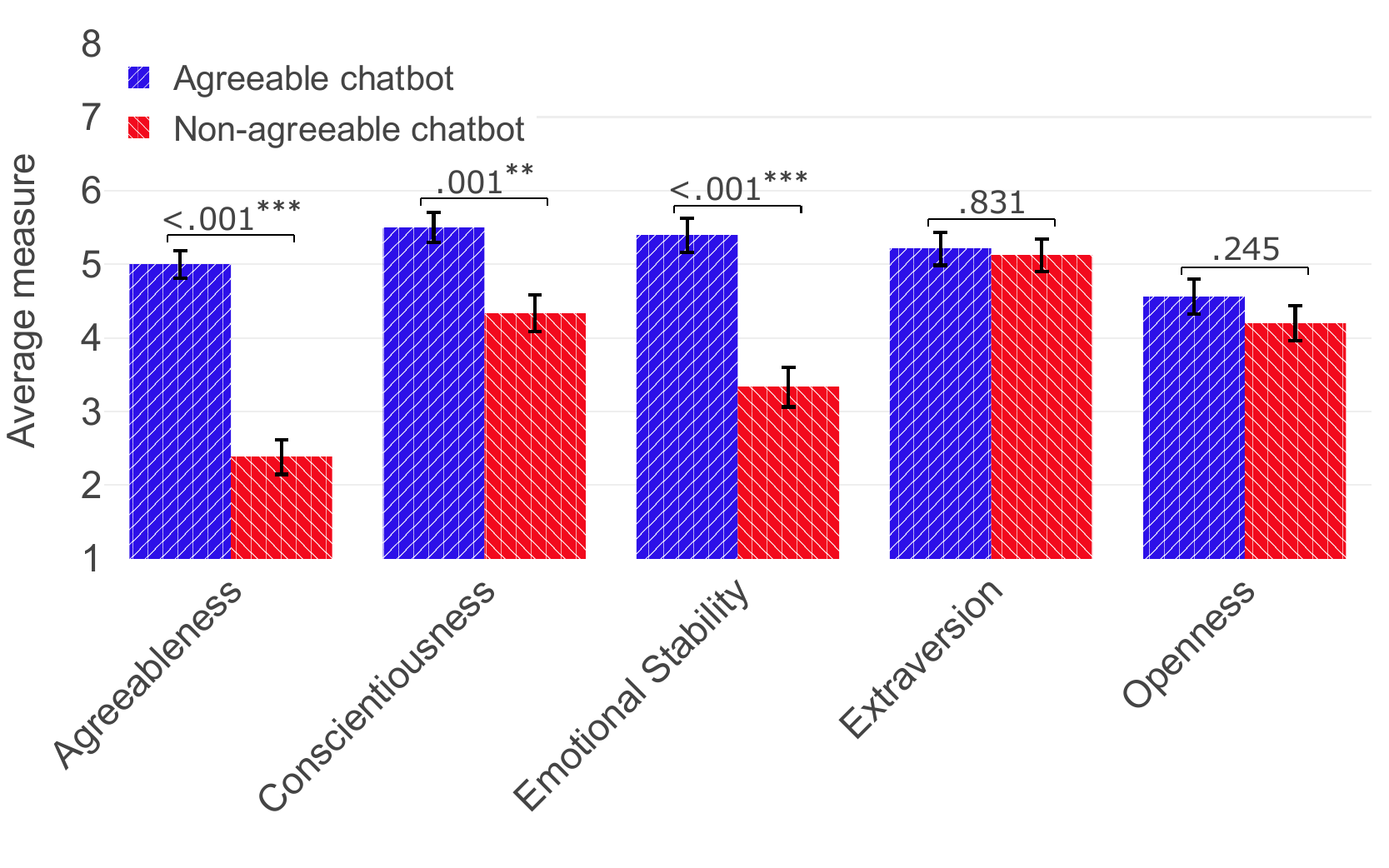}
    \caption{Assessed personality traits of both chatbots in the pre-study using TIPI. Values indicate \textit{p} value, \ensuremath{^{**}}\textit{p} $<$ .01, \ensuremath{^{***}}\textit{p} $<$ .001.}
    \label{fig:pre_robot_tipi}
\end{figure}

Both chatbots' personalities are assessed using the TIPI questionnaire, and the results are illustrated in Figure~\ref{fig:pre_robot_tipi}.
The Mann-Whitney U test shows that the agreeable chatbot ($W = 85.5$, $p <.001$) is perceived as more agreeable ($M = 5.00 $, $SD = 1.08 $) than the non-agreeable chatbot ($M = 2.38$, $SD = 1.34$). The agreeable chatbot is perceived as significantly ($W = 293$, $p = .001$) more conscientious ($M = 5.5$, $SD = 1.16$) than the non-agreeable chatbot ($M = 4.33$, $SD = 1.42$).
Additionally, the agreeable chatbot ($M = 5.39 $, $SD = 1.33 $) is perceived as more emotionally stable ($W = 171.5$, $p <.001$) than the non-agreeable chatbot ($M = 3.33$, $SD = 1.55$).

\subsection{Main Study Results}

\begin{figure}[h!]
\centering
    \includegraphics[width=0.49\textwidth]{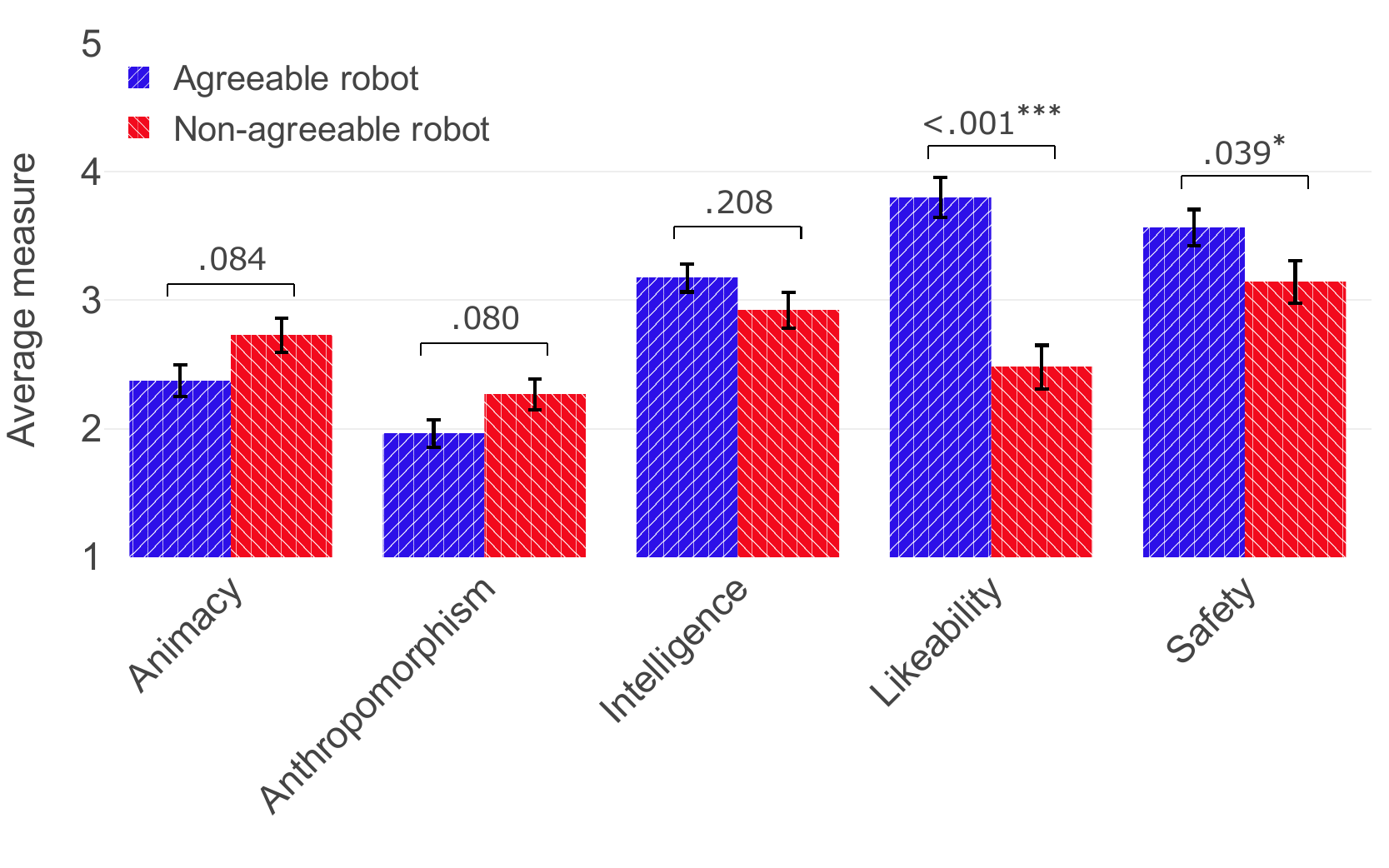}
    \caption{Godspeed questionnaire for the agreeable and non-agreeable robot in the main study. Values indicate \textit{p} value, \ensuremath{^{*}}\textit{p} $<$ .05, \ensuremath{^{***}}\textit{p} $<$ .001.}
    \label{fig:study_godspeed}
\end{figure}

Since the difference in agreeableness is evident in the pre-study, the implemented personalities are utilized for the robot in the main study. The difference in the perception of the robot, which is measured with the Godspeed questionnaire, is shown in Figure~\ref{fig:study_godspeed}.

The Mann-Whitney U test suggests a significant difference ($W = 837$, $p <.001$) in the robots' likeability between the agreeable ($M = 3.8$, $SD = 0.86$) and non-agreeable robot ($M = 2.48$, $SD = 0.98$). Furthermore, a significant difference ($W = 644.5$, $p = .039$) in the robots' perceived safety between the agreeable ($M = 3.57$, $SD = 0.77$) and non-agreeable robot ($M = 3.14$, $SD = 0.94$) is indicated.

\begin{figure}[h!]
\centering
            \includegraphics[width=0.49\textwidth]{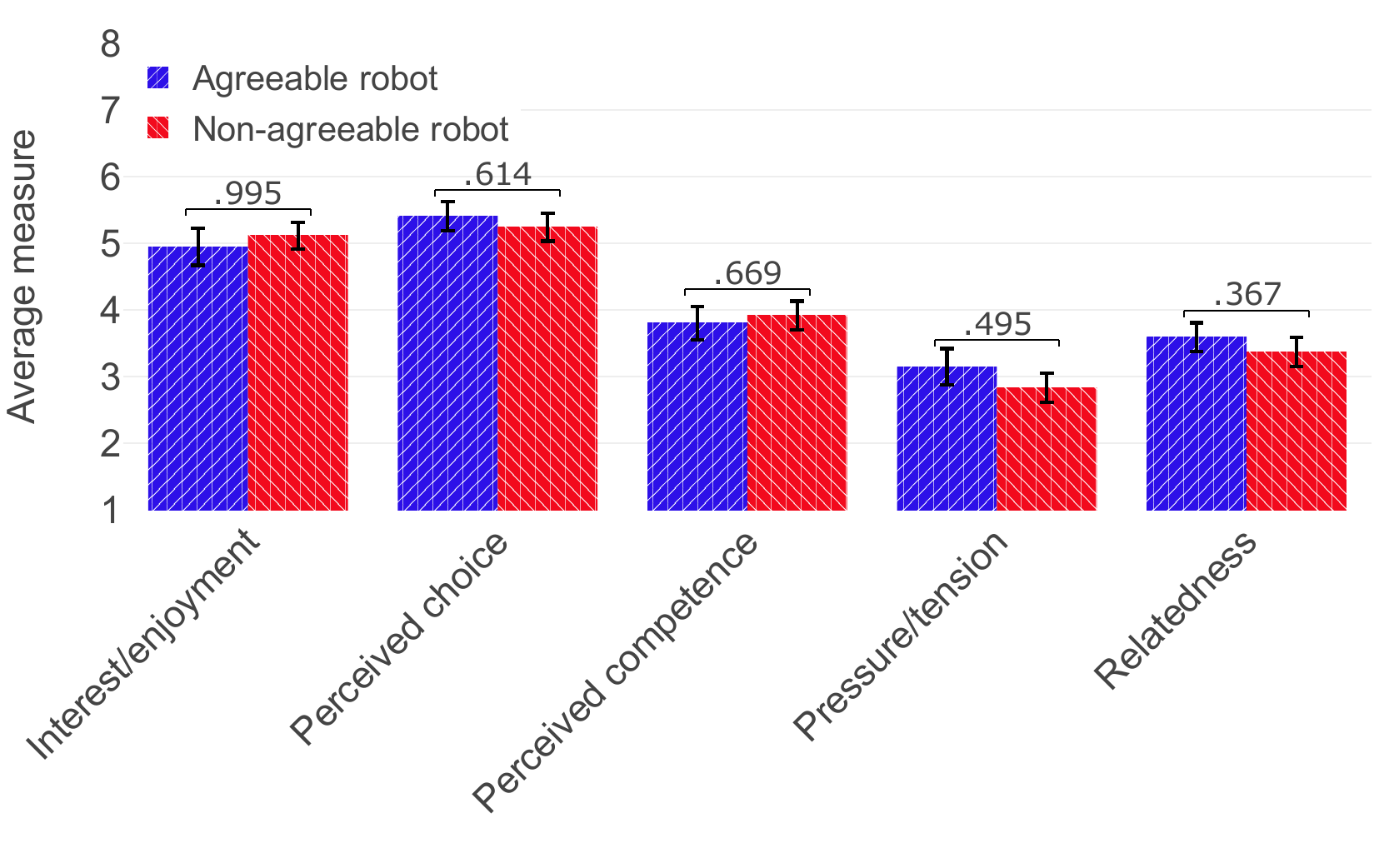}
    \caption{ IMI questionnaire for the agreeable and non-agreeable robot in the main study.}
    \label{fig:imi}
\end{figure}

As a measure of intrinsic motivation, the IMI questionnaire is assessed and is illustrated in Figure~\ref{fig:imi}. However, the Mann-Whitney U test does not reveal a significant difference between both experiment groups.

\begin{figure}[h!]
\centering
            \includegraphics[width=0.49\textwidth]{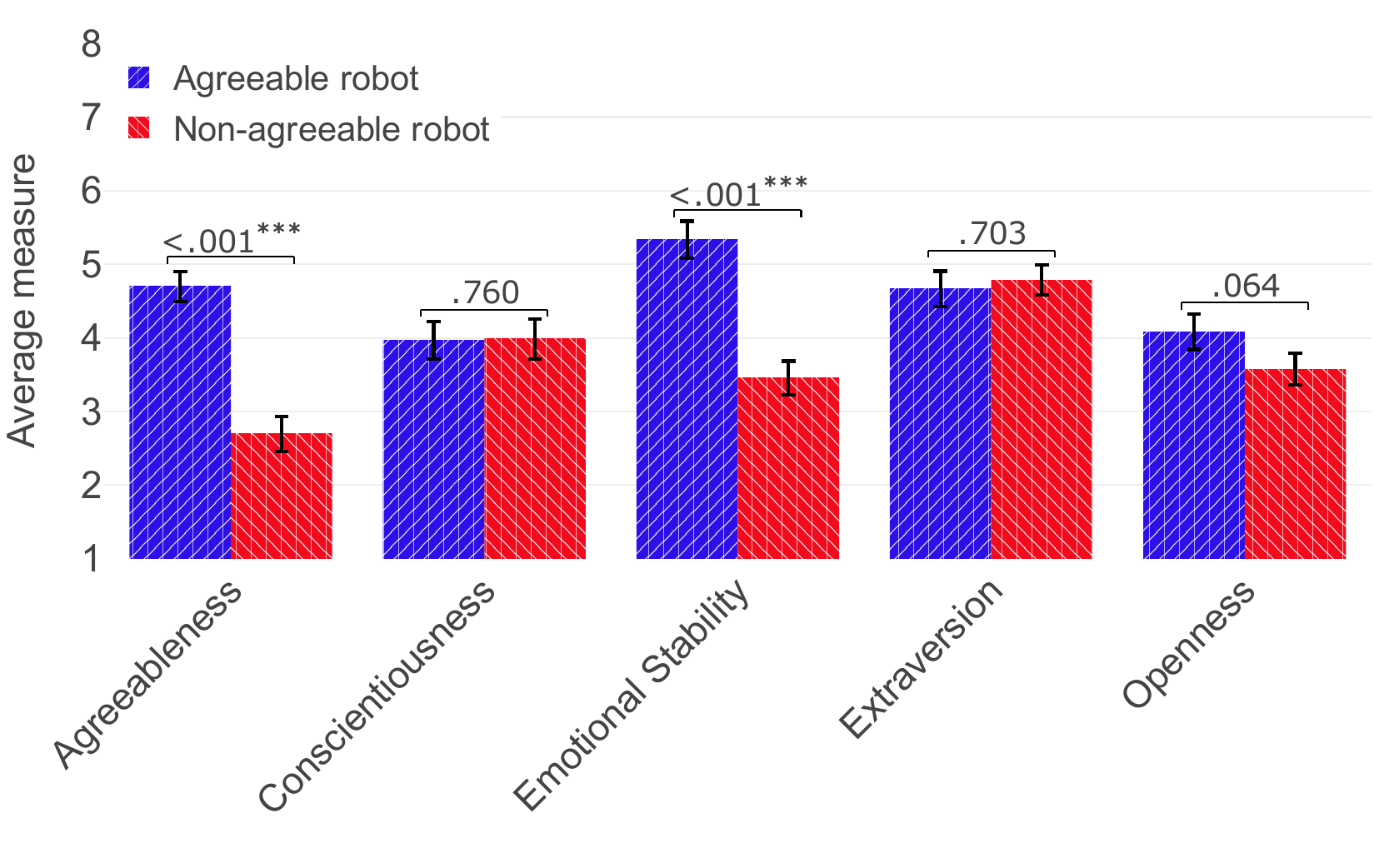}
    \caption{TIPI questionnaire for the agreeable and non-agreeable robot in the main study. Values indicate \textit{p} value, \ensuremath{^{***}}\textit{p} $<$ .001.}
    \label{fig:robot_tipi}
\end{figure}

Further, the robot's perceived personality is assessed using the TIPI questionnaire and is illustrated in Figure~\ref{fig:robot_tipi}. 
The Mann-Whitney U test shows a significant difference ($W = 860.5$, $p <.001$) in the robots' agreeableness between the agreeable ($M = 4.7$, $SD = 1.36$) and non-agreeable robot ($M = 2.70$, $SD = 1.36$). Furthermore, a significant difference ($W = 644.5$, $p <.001$) in the robots' emotional stability between the agreeable ($M = 5.33$, $SD = 1.39$) and non-agreeable robot ($M = 3.45$, $SD = 1.32$) is shown.

In the experiment, the participants' performance is measured as their final score, and their willingness to interact with the robot is measured as their experiment duration.

The participants in the agreeable robot condition achieved, on average, a final score of 6.6 with a standard deviation of 2.21.
In the non-agreeable experiment condition, the participants achieved an average score of 6.39 with a standard deviation of 1.91.
The Mann-Whitney U test does not suggest a significant difference ($W = 541.5$, $p = .522$) in the overall score between both experiment conditions.

For the experiment, the participants spent, on average, 15 minutes and 31 seconds with the agreeable robot, with a standard deviation of 5 minutes and 24 seconds. For the non-agreeable robot, the participants spent, on average, 17 minutes and 9 seconds, with a deviation of 9 minutes and 4 seconds. The Mann-Whitney U test does not suggest a significant difference ($W = 441$, $p = .464$).

\bgroup
\def\arraystretch{0.9}
\begin{table*}[h!]
    \caption{Kendall rank correlation between the interaction time and the assessed measures for the participants' motivation and the perceived robot's personality. \ensuremath{^{**}} denotes $p < .01$.}
\centering
\resizebox{\textwidth}{!}{\begin{tabular}{c|lllll|lllll}
\hline
\multicolumn{1}{c|}{} &\multicolumn{5}{c|}{IMI} & \multicolumn{5}{c}{TIPI}  \\

\hline
Item & Interest & Competence & Choice & Pressure & Relatedness & Agreeableness & Extraversion & Conscientiousness & Emotional Stability & Openness \\ 
$\tau$ & .255 & -.019 & .05 & -.018 & .139 & -.083 & .101 & .111 & -.009 & -.106  \\ 
 \textit{p} value & .004\ensuremath{^{**}} & \phantom{-}.826 & .575 & \phantom{-}.835 & .111 & \phantom{-}.355 & .263 & .221 & \phantom{-}.924 & \phantom{-}.24\\ 

   \hline
\end{tabular}}

    \label{tab:corr_exptime}
\end{table*}

\bgroup
\def\arraystretch{0.9}
\begin{table*}[h!]
    \caption{Kendall rank correlation between the participants' score and the assessed measures for the participants' motivation and the perceived robot's personality. \ensuremath{^{*}} denotes $p < .05$,  \ensuremath{^{**}} denotes $p < .01$.}
\centering
\resizebox{\textwidth}{!}{\begin{tabular}{c|lllll|lllll}
 \hline
\multicolumn{1}{c|}{} &\multicolumn{5}{c|}{IMI} & \multicolumn{5}{c}{TIPI}  \\

\hline
Item & Interest & Competence & Choice & Pressure & Relatedness & Agreeableness & Extraversion & Conscientiousness & Emotional Stability & Openness \\ 
$\tau$ & .147 & .201 & .216 & -.116 & .146 & .248 & .068 & .038 & .037 & .199\\ 
 \textit{p} value & .111 & .03\ensuremath{^{*}} & .021\ensuremath{^{*}} & \phantom{-}.211 & .115 & .009\ensuremath{^{**}} & .475 & .689 & .699 & .037\ensuremath{^{*}} \\

 \hline
\end{tabular}}

    \label{tab:corr_expscore}
\end{table*}

For the analysis of the assessed questionnaires and their influence on the experiment duration and final score, the Kendall rank correlation was estimated. 
The relationship between the participants' motivation and their perception of the robot and interaction time is shown in Table~\ref{tab:corr_exptime}.
Notably, the participants' interest exhibits a significant positive correlation with the interaction time ($\tau = .255$, $p= .004$), where a larger interest results in a longer interaction with the robot.

Table~\ref{tab:corr_expscore} shows the correlation between the questionnaires and the experiment score. It is notable that the motivation and the perception of the robot exhibit a relationship to the participants' performance.
Specifically, the reported competence ($\tau = .201$, $p= .03$) and the perceived freedom of choice ($\tau = .216$, $p= .021$) are associated with an increase in the final score. Likewise, the perceived robot's agreeableness ($\tau = .248$, $p= .009$) and openness ($\tau = .199$, $p= .037$) are linked with the participants' task performance.

\section{Discussion}
\label{sec:discussion}

\subsection{Pre-study Discussion}

In the pre-study, the responses generated by the language model for both personalities were evaluated in an object-guessing scenario. The participants read the conversation, and their perception of the chatbot's personality was assessed.

The evaluation of the pre-study demonstrates that both personalities were correctly recognized. The agreeable personality is perceived as more empathetic than the non-agreeable. Due to the non-agreeable chatbot being less agreeable, uninterested in the task, and occasionally impolite, it suggests a low level of empathy to the participants. Further, the participants attribute the non-agreeable chatbot with a lower intelligence. Since the chatbot is indifferent, less cooperative, and hinders the objective of the task, this could result in the impression that the non-agreeable robot does not understand the objective of the cooperative task, and is perceived as less intelligent.
The results indicate that the participants would prefer an agreeable chatbot as their partner in the task. An agreeable partner who is cooperative and focused would allow better teamwork and ensure task completion.

The assessed personality questionnaire (TIPI) for the chatbot shows, that the difference in agreeableness is notable by the participants (addressing RQ1).
In addition, a significant difference in conscientiousness can be observed. Conscientiousness refers to the degree of motivation, goal-directed behavior, and reliability. The non-agreeable chatbot is obstructing progress and thus might be perceived as less goal-directed and unreliable to cooperate and complete the assigned task.
Further, the non-agreeable chatbot appears less emotionally stable. Due to the snarky personality of the non-agreeable chatbot, it is perceived as less calm and composed than the agreeable chatbot.
Controlling one's emotions and acting calmly might be preferred traits in a teammate and would enable steady progress in the task, even in the presence of pressure or misunderstandings.

\subsection{Main Study Discussion}
In the main study, the participants engaged with a robot in a game of Quickdraw, with either the agreeable or non-agreeable personality that was evaluated in the pre-study. The Godspeed questionnaire shows that the agreeable robot was perceived as more likable than the non-agreeable robot.

Since the agreeable robot was more cooperative and friendly, the participants liked the agreeable robot better. This strong link between agreeableness and likability is likewise suggested in the literature~\cite{Castro-Gonzalez2016}.
Although the physical appearance of the robot and the environment in both experiment conditions are identical, the agreeable robot is considered safer. This perceived difference in safety might be attributed to the non-agreeable robot's personality appearing more threatening and less collected and calm. Regarding RQ2, the results show that a robot's agreeableness increases its likability and suggest that other personality traits can be affected by a robot's agreeable personality.

Despite the noticeable difference in both robots' personalities, the IMI questionnaire does not provide evidence of an effect on the participants' motivation (addressing RQ3). Despite the non-agreeable robot being rude, it did not affect the participants' motivation. Since the participants were aware that they were part of an experiment and the cooperation partner was a robot, it might be easier to brush off snarky comments, which could be interpreted as the robot's sense of humor, and instead, the participants focused on the drawing task. Furthermore, some participants might perceive the non-agreeable robot as more interesting.
Participants might have different preferences for the robot's personality to encourage them to stay motivated. 
Previous research suggests that matching a participant's personality with the robot's personality can positively affect motivation~\cite{Andrist2015b}.

The assessed TIPI questionnaire is in accordance with the pre-study, and shows a difference in agreeableness. This confirms that the personality can be observed in either an online study or in a laboratory experiment with an embodied robot. Further, the difference in emotional stability is also present where the manipulation of the agreeableness personality trait affects perceived emotional stability. However, in contrast to the online pre-study, there is no measurable difference in the robot's conscientiousness. Suggesting that the robot in the main study, despite its different personalities, was perceived as similar in motivation and goal-directed behavior. With respect to RQ1, an LLM can be utilized for a robot to consistently portray a specific personality, as shown by both studies.

The experiment time and final score do not reveal a difference between both conditions.
Besides the interaction during drawing, extended verbal interaction with the robot is not mandatory in the experiment. Therefore, the participants in both groups might not have felt the need to converse with the robot and instead focused on completing the task.
Although research suggests that emotional stability, a relaxed atmosphere, and agreeableness foster a cooperative environment, and can improve task performance~\cite{Peeters2006}, there is no difference in scores between both experiments. With respect to RQ3, we do not observe a direct effect on motivation and task performance. The complexity of the experiment, quickly sketching an image on the provided trackpad, might be challenging and could require multiple sessions to measure an improvement in skill and motivation.

The correlation analysis reveals that interest is positively correlated with interaction time. A person with a greater interest in the task and the robot might explore the experiment and engage in more conversation with the robot.
The participant's competence correlates with the experiment score. People with drawing experience perceive themselves as more competent and are likely to score higher. Likewise, the perceived freedom of choice in the experiment can increase the obtained score. 
The participants might feel more relaxed and have the freedom to adjust their sketching style to improve the recognizability of their drawing for the robot. 

Regarding the perceived robot's personality, the robot's agreeableness and openness can improve the final score.
Despite no evident difference in the participants' scores between both experiment groups, it is suggested that the perceived agreeableness and openness can increase the task performance for some participants. An agreeable robot might support and encourage the participants, as observed in the pre-study, where an agreeable robot is preferred as a teammate. Even if a participant fails in one round, the robot's positive attitude and perceived engagement in the task can be encouraging. Likewise, the perceived openness of the robot might encourage the participants to explore different methods to ensure success in the next round. In contrast, lower openness and acceptance will discourage the participants from trying new approaches to solve the task.


\section{Limitations}
\label{sec:limitations}

Despite the promising results, several avenues for future research should be explored. This study focused primarily on the personality trait of agreeableness, but the effects of other personality traits in cooperative tasks warrant further investigation. Since changes in one personality trait can influence other traits, as demonstrated in this study, it is essential to examine the correlations and interplay between multiple traits in human-robot interaction. Although this research suggests that an agreeable personality is generally favorable, there may be applications where a non-agreeable personality can be beneficial. 
In addition, this study examined a cooperative task involving one human and one robot. Future research could explore scenarios involving additional actors, where both the human and the robot collaborate against others. Such situations may amplify the observed effects and provide deeper insights into the complexity of multi-agent cooperation.
Finally, to better combine the modalities of vision and language, a vision-language model could be used to obtain a better understanding of what the participant is drawing and improve the accuracy of the robot's comments.


\section{Conclusion}
\label{sec:conclusion}

This study investigated the implementation of agreeable and non-agreeable personalities on a robot using an LLM and assessed their impact on a cooperative human-robot task. The research consists of two parts: an online pre-study and a lab-based main study, where participants engaged with the robot in a game of Quickdraw.

The pre-study confirmed that the manipulation of agreeableness in the LLM's personality was consistent. Participants expressed in the prestudy's questionnaire a preference for an agreeable partner in cooperative tasks. The main experiment shows a strong relationship between the robot's agreeableness and its likability, suggesting that agreeable robots are generally more favored by participants. However, there is no direct evidence linking the robot's personality to differences in motivation or task performance. However, a correlation was observed between scores obtained by participants and their perception of the robot's openness to new experiences and agreeableness, implying that such traits may encourage participants and potentially enhance task performance.
Factors that influence a participant's motivation might be individual and task-dependent. While some participants could prefer an encouraging and agreeable robot, others could prefer a snarky robot that teases them to improve their skills. Thus, matching an individual preference with the robot's personality would be beneficial. Furthermore, a single interaction with the robot and the task might be insufficient to measure changes in motivation and task performance. Repeated interaction might be required to measure an improvement in task performance, and willingness to return and interact with the robot could indicate motivation.

Overall, this study highlights the utility of LLMs in endowing robots with consistent personalities and provides valuable insights into the influence of personality on human-robot collaboration. In particular, it supports the link between agreeableness and likability and suggests that agreeable robots may improve performance in cooperative tasks for some participants. Our research opens new avenues towards investigating the complex effects of robot personality and behavior, particularly with respect to motivation and intention to use, which may become more evident with repeated interactions.


\begin{acks}
Many thanks to Alhassan Abdelhalim, Emad Aghajanzadeh, Zeynep Seda Birinci, David Fidanyan, Aron Teodor Jinga, Pascal Lindner, Aditi Mhatre, Atharva Yashodhan Phatak, Bishab Pokharel, Vidushi Rawat, Aimen Shahid, Archit Sharma, Ismail Barkin Ulusoy, and Aamir Waris for their contribution to the project.
The authors gratefully acknowledge support from the German Research Foundation DFG (CML, LeCAREbot), the European Commission (TRAIL, TERAIS), and the Federal Ministry for Economic Affairs and Climate Action (BMWK) under the Federal Aviation Research Programme (LuFO), Projekt VeriKAS. 
\end{acks}

\appendix
 \section{Chatbot perception questionnaire} \label{Appendix:chatbotQustionaire}

\begin{itemize}

    \item How relevant and appropriate were the responses of the chatbot? 
    <\textit{Not relevant at all} | \textit{Highly relevant}>

    \item Did you feel the chatbot responded empathetically?
    <\textit{Did not empathize at all} |  \textit{Fully empathized}>
    
    \item How intelligent did you perceive the chatbot to be?
    <\textit{Not at all intelligent} | \textit{Very intelligent}>
    
    \item How easy did you perceive the interaction with the chatbot to be?
    <\textit{Very difficult} | \textit{Very easy}>
    
    \item How likely would you like to have this chatbot as a teammate?
    <\textit{Very unlikely} | \textit{Very Likely}>
    
    \item Did you feel the chatbot understood the queries?
    <\textit{Did not understand at all} | \textit{Fully understood}>
    
\end{itemize}

  \bibliographystyle{ACM-Reference-Format}
  \bibliography{main}



\end{document}